\documentstyle[12pt]{article}

\def\lam{\lambda}

\def\del{\partial}
\def\nab{\nabla}
\def\til{\tilde}
\def\dis{\displaystyle}

\def\a{\rm a}
\def\b{\rm b}

\begin{document}

\begin{center}
{\large\bf Interpretation of the first order formalism of $f(R)$-type gravity
and the corresponding second order formalism}
\end{center}
\hspace*{15mm}\begin{minipage}{13.5cm}
Y. Ezawa, H. Iwasaki, Y. Ohkuwa$^{\dagger}$, T. Uegaki, N. Yamada and 
T. Yano$^{*}$\\[3mm]
Department of Physics, Ehime University, Matsuyama, 790-8577, Japan\\[1mm]
\hspace*{-1.5mm}$^{\dagger}$Department of Mathematics, Miyazaki Medical 
College, University of Miyazaki, Kiyotake, Miyazaki,  889-1692, Japan\\[1mm]
\hspace*{-1.5mm}$^{*}$Department of Electrical Engineering, Ehime University, 
Matsuyama, 790-\\[.5mm]
8577, Japan\\[5mm]
{\small Email :  ezawa@sci.ehime-u.ac.jp, hirofumi@phys.sci.ehime-u.ac.jp,\\
ohkuwa@med.miyazaki-u.ac.jp, naohito@phys.sci.ehime-u.ac.jp\\
 and yanota@eng.ehime-u.ac.jp}
\\[8mm]
{\bf Abstract}\\[2mm]
 $f(R)$-type gravity in the first order formalism is interpreted as Einstein 
gravity with non-minimal coupling arising from the use of unphysical frame.
Identification of the corresponding second order higher-curvature gravity in 
the physical frame is proposed by requiring that the action is the same.

\end{minipage}

\section{Introduction}

Up to the present, our knowledge of the universe has become wider both 
observationally and theoretically.
So the theories of gravity, the dominant force in the universe, have been 
extended and variety of generalized theories of gravity have been proposed, 
starting from the Einstein one where the Lagrangian density is a linear 
function of the scalar curvature.
The natural motivation arises when we consider the initial singularity of our 
universe\cite{EH,Wald} where the gravity should be so strong that higher 
powers of the invariants constructed from the scalar curvature, the Ricci 
tensor and the Riemann tensor, would play important roles\cite{N-NT-TAN}.
Other examples are given when we consider the quantum field theory in a curved 
spacetime \cite{BD}, the effects of gravity on strings, extended objects, 
since they recieve the tidal force described by the curvature.
The local expression of the latter leads to the modification of gravity by 
higher powers of the curvature tensors.
The string perturbation theory, in fact, gives a definite combination known as 
the Gauss-Bonnet combination\cite{Zwie}.

Recent observations of the distant type Ia supernonae suggest that 
the universe is presently in the period of accelerating expansion
\cite{SNIa,CMB}.
There have been two ways to explain this.
One is to introduce new kind of gravitational sources such as dark energy.
The other is to modify the law of gravity.
The latter includes the introduction of a term proportional to the inverse 
power of the scalar curvature\cite{CCTetc,VMF}.
So, in this work, we confine ourselves to the theories where the Lagrangian 
density is given by a function of the scalar curvature $f(R)$.
This type of theories have long been investigated and expected to resolve 
the problem of initial singularity\cite{N-NT-TAN,S-S,BHV,EIOUYY}.
Thus they are a natural and convenient generalization of Einstein gravity both 
to strong and weak gravity regions.

Theoretically conformal equivalence of this theory to the Einstein one are 
well known\cite{MS,CMQ}.
However, identification of the physical metric has not been settled.
Other theoretical problem seems to be a more fundamental one, i.e. how to 
apply the variational principle.
It is known that the first order, or Palatini, formalism of Einstein gravity 
is equivalent to the second order formalism.
We mean by the second order formalism the one where the connection is given by 
the Christoffel symbol, so that the metric tensor is the only variable 
describing gravity.
In this sence, the first order formalism is another line of generalizing 
Einstein gravity, since the connection is not given {\it a priori}, but is 
determined by the physical law, the variational principle.
The above equivalence means that the relation of the connection to 
the Christoffel symbol determined by the variational principle is that they 
are the same.
However, for general $f(R)$-type gravity, they are different.
This was noticed from the fact that the connection is not the Christoffel 
symbol but is given by e.g. the relation (5) below\cite{S-S,BHV,VMF,CMQ}.
Recently Flanagan showed that this difference is also seen by transforming to 
the conformally related Einstein frame where the scalar field has no 
kinematical term contrary to the  second order formalism\cite{VMF}.
Here we will present another way of expressing the difference by clarifying 
the correspondence of the first order formalism and the second order formalism
by constructing the Lagrangian density to be used in the second order 
formalism starting from the one in the first order formalism.
In doing so we use the fact that the connection in the first order formalism 
is just the conformal transformation of the Christoffel symbol as seen in (5a).

\section{Interpretation of the first order formalism}

We consider a theory described by an action
$$
S=\int \sqrt{-g}f(R)d^nx                                           \eqno(1)
$$
where
$$
R=g^{\mu\nu}R_{\mu\nu},\ \ \ g=\det g_{\mu\nu}.                    \eqno(2)
$$
$g_{\mu\nu}$ is the metric tensor and $R_{\mu\nu}$ the Ricci tensor.
In the first order formalism $R_{\mu\nu}$ is given by the usual expression 
in terms of the connection $\Gamma^{\lam}_{\mu\nu}$ taken as independent 
variables.
This formalism is in line with the genaralization of gravity since the metric 
and the connection are independent concepts geometrically.
Field equations in the first order formalism are
$$
\left\{\begin{array}{l}
\dis f'(R)R^{\mu\nu}-{1\over2}f(R)g^{\mu\nu}
={1\over2}T^{\mu\nu}\\[5mm]
\dis {\nab}_{\lam}\left\{\sqrt{-g}g^{\mu\nu}f'(R)\right\}=0.
\end{array}\right.                                                 \eqno(3)
$$
By taking the trace of the first equation and solving it for $R$, we have
$$
R=R(T).                                                            \eqno(4)
$$
where $T$ is the trace of $T_{\mu\nu}$.
This means that $R$ and consequently $f'(R)$ is expressed by the metric 
and the matter variables.

It is well known that the second equation in (3) is rewritten as
$$
\Gamma^{\lam}_{\mu\nu}=\Gamma^{(0)\lam}_{\mu\nu}+{1\over n-2}\left[
\delta^{\lam}_{\nu}\del_{\mu}(\ln f')
+\delta^{\lam}_{\mu}\del_{\nu}(\ln f')
-g_{\mu\nu}\del^{\lam}(\ln f')\right]                              \eqno(5\a)
$$
or
$$
\Gamma^{\lam}_{\mu\nu}
=\Gamma^{\lam(0)}_{\mu\nu}
+{f''(R)\over (n-2)f'(R)}\left[\delta^{\lam}_{\mu}\del_{\nu}R
+\delta^{\lam}_{\nu}\del_{\mu}R-g_{\mu\nu}\del^{\lam}R\right]      \eqno(5\b)
$$
where $^{(0)}$ denotes that the quantity is constructed from the metric
(in this case the Christoffel symbol).
This means with (4) that the connection, though treated as independent 
variables is expressed solely by the metric and the matter variables.

We are left with determining the metric (and the matter variables).
The equations for the metric is obtained as
$$
G^{(0)}_{\mu\nu}+\Lambda(T)g_{\mu\nu}={\til{T}_{\mu\nu}\over 2f'}  \eqno(6)
$$
by using (5) in the first equation of (3).
Here
$$
\Lambda(T)\equiv 
{1\over2}\left[R(T)+{(n-2)(n-3)\over4}b_{\lam}b^{\lam}
                   -{f\over f'}\right]                             \eqno(7)
$$
and
$$
\til{T}_{\mu\nu}=T_{\mu\nu}+(n-2)f'\left[\nab_{\mu}b_{\nu}
     -\nab_{\lam}b^{\lam}g_{\mu\nu}-{1\over 2}b_{\mu}b_{\nu}\right]\eqno(8)
$$
where
$$
b_{\mu}\equiv {2\over n-2}\del_{\mu}(\ln f')                       \eqno(9)
$$
Thus the $f(R)$-type gravity in the first order formalism should not be 
interpreted as a higher-curvature gravity but an  gEinstein gravity"
coupled non-minimally with matter.
This interpretation is consistent with the Flanagan's result mentioned above
\cite{VMF},@indicating that the dynamical degrees of freedom of gravity are 
described by the metric tensor alone.
The non-minimality of the coupling may be related to the criteria for physical 
metric by Magnano and Sokolowski that the coupling should be minimal in 
the physical frame\cite{MS}.

The situation is made clearer by examining approximate solutions obtained by 
taking deviations from the Einstein one as perturbations.
Consider the case where $f(R)$ is quadratic in $R$
$$
f(R)={1\over 2\kappa}(-2\Lambda +R+\xi R^2).                       \eqno(10)
$$
We apply this model to the flat Robertson-Walker spacetime filled with perfect 
fluid and examine the effects of higher-curvature term on the evolution of 
the scale factor $a(t)$.(We take $n=4$ for a while.)
Expanding $a(t)$ as
$$
a(t)=a^{(0)}(t)+\xi a^{(1)}(t),                                    \eqno(11)
$$
where $a^{(0)}(t)$ is the scale factor in Einstein gravity.
Retaining the linear terms in $\xi$ and assuming the matter dominance and 
putting $\Lambda=0$, we have
$$
a(t)=a^{(0)}(t)\left[1+{5\til{\xi}\beta^2\over 16(1-3\beta/8)}\left\{
(t/t_{0})^{-2}-(t/t_{0})^{-3\beta/4}\right\}\right]                \eqno(12)
$$
for the first order formalism and
$$
a(t)=a^{(0)}(t)\left[1-{3\til{\xi}\over 1-3\beta/8}\left\{
(t/t_{0})^{-2}-(t/t_{0})^{-3\beta/4}\right\}\right]                \eqno(13)
$$
for the second order formalism.
Here $\beta\equiv \kappa\rho_{0}t_{0}^2\approx 1.33$ and 
$\til{\xi}\equiv \xi/t^2_{0}$ and subscript 0 means the present value (or 
strictly, a certain time in the matter dominant era).
It is seen that, in the first order formalism, the deviation from the Einstein 
gravity is proportional to $\beta^2$ representing the nonminimality.
It is noted that the Jeans mass, the critical mass in the structure formation, 
is proportional to $a^{-3/2}(t)$, so that it may be possible to detect 
the effect of higher-curvature term observationally.

\section{Correspondence}

Equation (6) is the one in Einstein gravity coupled non-minimally.
As noted above, this could be interpreted that the frame is unphysical.
Corresponding physical frame is suggested by (5a,b) which represent 
the conformal transformation of the Christoffel symbol, i.e. the connection 
$\Gamma^{\lam}_{\mu\nu}$ is the conformal transformation of the Christoffel 
symbol, $\Gamma^{(0)\lam}_{\mu\nu}$.
The transformation is as follows
$$
\bar{g}_{\mu\nu}=\Omega^2g_{\mu\nu},\ \ \ 
\Omega^2=\left[k\times f'(R)\right]^{2\over n-2},\ (k=constant).   \eqno(14)
$$
The constant $k$ is chosen such that $\Omega=1$ for Einstein gravity. 
Then the relation (5a,b) is written as:
$$
\Gamma^{\lam}_{\mu\nu}=\bar{\Gamma}^{\lam(0)}_{\mu\nu}             \eqno(15)
$$
This means that the theory becomes the second order if $\bar{g}_{\mu\nu}$ is 
used as the metric tensor.
The transition is straightforward as is shown below.
Thus introduction of the connection as independent variables to describe 
gravity seems to be redundant.
The action in the second order formalism is taken to be the action (1) 
rewriten in terms of the metric $\bar{g}_{\mu\nu}$.
Equation (15) means 
$$
R_{\mu\nu}=\bar{R}_{\mu\nu}^{(0)}
$$
which in turn means
$$
R=\left[kf'(R)\right]^{2\over n-2}\bar{R}^{(0)}.                   \eqno(16)
$$
Note that the contraction is taken between the untransformed metric and 
the transformed Ricci tensor.
Solving the equation (16) with respect to $R$, we can write
$$
R=\zeta(\bar{R}^{(0)}).                                            \eqno(17)
$$
In conformally transformed frame, only metric compatible quantities will be 
used, so we drop the superscript $^{(0)}$ in this frame hereafter.
Noting that $g=\bar{g}/[kf'(R)]^{2n\over n-2}
=\bar{g}/[kf'(\zeta(\bar{R}))]^{2n\over n-2}$, we can rewrite (1) as
$$
S=\int\sqrt{-\bar{g}}F(\bar{R})d^nx                                \eqno(18)
$$
where
$$
F(\bar{R})
={f\left(\zeta(\bar{R})\right)\over 
\left[kf'\left(\zeta(\bar{R})\right)\right]^{n\over n-2}}          \eqno(19\a)
$$
which is also expressed, in a more symmetrical form, as
$$
{F(\bar{R})\over \bar{R}^{n\over 2}}={f(R)\over R^{n\over2}}       \eqno(19\b)
$$
Equation (18) gives the action to be used in the second order formalism 
corresponding  to the action (1) to be used in the first order formalism.\\
The functional form of $f(R)$ and $F(\bar{R})$ are different in general.
This is another way of indicating that the second order formalism, in which 
the action (1) is taken as $f(R^{(0)})$, would be different from the one in 
the first order formalism.
We will refer the frame in which the metric is $\bar{g}_{\mu\nu}$ as 
the second order frame.

Finally we will illustrate the correspondence by examples for $f(R)$:\\
(i) $f(R)=a+bR+cR^2$

Equation (16) reads in this case as
$$
R=\biggl[\frac{1}{b}(b+2cR) \biggr]^\frac{2}{n-2}\bar{R}           \eqno(20)
$$
where $k$ is chosen to satisfy $kb=1$ which leads to $\bar{R}=R$ for $c=0$.
It is difficult to solve this equation for arbitrary $n$.
For $n=4$, $F(\bar{R})$ is given by
$$
F(\bar{R})=a+\left(1-{4ac\over b^2}\right)(b\bar{R}-c\bar{R}^2).   \eqno(21)
$$
We can also solve for $n=3$.\\
(ii) $f(R)=a+bR+cR^{n\over2}$\\
The relation between $R$ and $\bar{R}$ is given by
$$
R=\frac{\bar{R}}{\left(1-\frac{nc}{2b}\bar{R}^\frac{n-2}{2}
\right)^\frac{2}{n-2}} \ , 
$$
where $k$ is chosen to satisfy $kb=1$ again and $F(\bar{R})$ is expressed as
$$
F(\bar{R})= a \left(1-\frac{nc}{2b}\bar{R}^\frac{n-2}{2}
\right)^\frac{n}{n-2}
+b\bar{R}+c\left( 1-\frac{n}{2}  \right) \bar{R}^\frac{n}{2} \ .   \eqno(22)
$$
It is seen that (22) reduces to (21) for $n=4$.
For other cases it is difficult to solve (16) as in the example (i).
For $n=4$, (16) can be solved for the following examples:\\
(iii) $f(R)=a+bR+cR^2+dR^3$
$$
\begin{array}{lcl}
F(\bar{R})&=&\dis {1\over2}a+{2\over3}b\left(1-{3ac\over b^2}\right)\bar{R}
           -{c\over3}\left(1-{6ac\over b^2}+{9ad\over bc}\right)\bar{R}^2
\\[5mm]
&\pm&\dis\left[{1\over2}a+{1\over3}b\left(1-{3ac\over b^2}\right)\bar{R}\right]
\sqrt{\left(1-{2c\over b}\bar{R}\right)^2-{12d\over b}\bar{R}^2}.
\end{array}                                                        \eqno(23)
$$
(iv) $f(R)=a+bR+c/R$, equation (16) becomes a cubic equation
$$
R^3-\bar{R}R^2+(c/b)\bar{R}=0                                      \eqno(24)
$$
where we put $kb=1$. 
The solutions of this equation are known to be given by a set of formulae 
which are not cited here.
We only note that with familiar notations, $b=1/2\kappa$ and 
$c=-\mu^4/2\kappa$, (24) has a real solution 
$$
R=(\alpha)^{1/3}+(\beta)^{1/3}+\bar{R}/3                           \eqno(25)
$$
where
$$
\alpha,\beta
=(\bar{R}/3)^3+{\mu^4\bar{R}\over 2}\left(
                       1\pm \sqrt{1+4\bar{R}^2/27\mu^4}\right)     \eqno(26)
$$
Thus for this type of $f(R)$, although the second order form could also be 
described by $f(R^{(0)})$, the results in general differ qualitatively from 
the case when $f(R)$ is used in the first order formalism.
However, when $\mu^4/R$ term is small, i.e. the theory is approximately 
Einstein, $F(\bar{R})\approx a+(\bar{R}+2\mu^4/\bar{R})/2\kappa$ as expected.

\section{Summary and discussion}
 
We argued that gravity described by the Lagrangian density $\sqrt{-g}f(R)$ in 
the first order formalism should be interpreted as Einstein one.
The reasons are that the connection is expressed solely in terms of the metric 
and the matter variables and that, in the conformally transformed Einstein 
frame, the scalar field has no dynamical degrees of freedom.
We also argued that the frame in which the Lagrangian density takes the above 
form is unphysical.
The conformal transformation to the physical frame is identified and 
the Lagrangian density in the physical frame, $\sqrt{-\bar{g}}F(\bar{R})$, is 
obtained to be used in the second order formalism.
The functional form of $f(R)$ and $F(\bar{R})$ are different except for 
the Einstein case.
It is noted, however, for quadratic $f(R)$, $F(\bar{R})$ is also quadratic 
for the spacetime dimension $n=4$, although the coefficients are different.

The Lagrangian density $\sqrt{-\bar{g}}F(\bar{R})$ may in turn be conformally 
transformed, so that it describe Einstein gravity with a scalar field.
We would like to comment on this conformal equivalence.
If we take the relevant conformal transformation to be a change of variables, 
equivalence is clear except that which metric is physical, i.e. should be 
identified as, e.g. FRW metric.
However, when the dynamical aspects are concerned, situation is not so clear 
because the transformation depends on the curvature.
The curvature includes the second order derivative of the metric.
From the viewpoint of canonical formalism, the transformation mixes 
the coordinates with the momenta, i.e. the transformation is the one in 
the phase space.
Therefore, unless the transformation turns out to be a canonical one, 
the dynamical system is changed by the transformation.
The canonical transformation of the higher-curvature gravity has some subtle 
points.
The well known Ostrogradski's method\cite{Ostro} is not directly applicable 
since the curvature depends on the time derivatives of the lapse function and 
the shift vector.
The generalization was made by Buchbinder and Lyakhovich\cite{BL} and applied 
to the higher-curvature gravity theories.
However in their method, it seems to be inevitable that changes of 
the generalized coordinates lead to the change of the Hamiltonian even if 
the changes are independent of time.

Thus, concerning the physical nature of the metric, a new problem arises 
whether the metric $\bar{g}_{\mu\nu}$ in the second order frame is physical or 
not, although we here assumed the physical frame to be the one where 
the connection is the Christoffel symbol.
These problems will be investigated in separate works.

\end{document}